\documentclass[12pt]{article}
\usepackage{amsmath}
\usepackage{graphicx}
\usepackage{amsfonts}
\usepackage{amssymb}

\newdimen\dummy
\dummy=\oddsidemargin
\begin{document}
\pagestyle{empty}

\title {DARK MATTER AND SCALE INVARIANCE}
\author{Hung Cheng\\Department of Mathematics\\Massachusetts
  Institute of Technology\\Cambridge, MA  02139}

\date{}

\maketitle

\begin{center}

\begin{abstract}

We point out that the vector meson $S$ first proposed by Weyl on the basis
of
scale invariance is a candidate for dark matter.

\end{abstract}

\end{center}

\bigskip

As colliders are being built with the creation of dark matters among the
stated goals, we wish to point out that one of the possible candidates of
dark
matter is the vector meson $S$ first conceived by Weyl on the basis of
scale
invariance$^{1,2}$. \

Weyl made the mistake of identifying $S$ with the photon. Also, Weyl's
work is
incomplete, as he lacked at his time the framework of the theory of
quantum
electrodynamics. Our formulation of scale invariance yields results$^{3}$
different from those of Weyl's. For example, we find that $S$ does not
couple
with the electron. It is well-known that Einstein argued that electrons
moving
in a background of the $S$ field would produce unobserved broading of the
atomic spectral lines. That the electron does not interact with $S$
overrides
this objection of Einstein. Since Weyl's theory was first proposed in
1919,
before even the discovery of Dirac's equation, Weyl did not have the
prerequisites which might have led him to conclude that $S$ is a kind of
dark
matter coupling with neither quarks and leptons, nor any of the gauge
mesons.
Indeed, in our theory, the only interactions $S$ has are those with the
graviton and scalar mesons, the only candidate of which at the moment is
the
Higgs meson.

Let $g_{\mu\nu}$ be the metric tensor. Then the distance between two
neighboring space-time points is

\bigskip

$ds^{2}=g_{\mu\nu}dx^{\mu}dx^{\nu}.$

\bigskip

Let us change the scale of distance globally, e.g., changing the unit of
length from the centimeter to the inch for every point in space. With this
change, the distance remains the same, but is measured in a different
unit.
This can be done by changing the metric tensor from $g_{\mu\nu}$ to
$g_{\mu
\nu}^{\prime},$ where

\bigskip

$g_{\mu\nu}^{\prime}=\Lambda^{2}g_{\mu\nu},$ $\ \ \ \ \ \ \ \ \ \ \ \ \ \
\ \ \ \ \ \ \ \ \ \ \ \ \ \ \ \ \ \ \ \ \ \ \ \ \ \ \ \ \ \ \ \ \ \ \ \ \
\ \ \ \ (1)$

\bigskip

with $\Lambda$ a constant. Then we have

$\bigskip$

$ds^{\prime2}=\Lambda^{2}ds^{2},$

\bigskip

where

$\bigskip$

$ds^{\prime2}\equiv g_{\mu\nu}^{\prime}dx^{\mu}dx^{\nu}.$

\bigskip

Thus, with the transformation (1), the numerical value of the distance
between
two given points change by a constant multiple. Since $g^{\mu\nu}$ is the
inverse of $g_{\mu\nu}$, we have from (1) that

\bigskip

$g^{\prime\mu\nu}=\Lambda^{-2}g^{\mu\nu}.$ \ \ \ \ \ \ \ \ \ \ \ \ \ \ \ \
\ \ \ \ \ \ \ \ \ \ \ \ \ \ \ \ \ \ \ \ \ \ \ \ \ \ \ \ \ \ \ \ \ \ \ \ \
\ \ \ \ \ \ \ \ \ \ \ \ \ \ \ \ \ \ \ \ (2)

\bigskip

We also have from (1) that

$\bigskip$

$\mid g^{\prime}\mid^{1/2}=\Lambda^{4}\mid g\mid^{1/2},\ $\ $\ \ \ \ \ \ \
\ \ \ \ \ \ \ \ \ \ \ \ \ \ \ \ \ \ \ \ \ \ \ \ \ \ \ \ \ \ \ \ \ \ \ \ \
\ \ \ \ \ \ \ \ \ \ (3)$

\bigskip

where $\mid g\mid$ is the determinant of the matrix $g_{\mu\nu}.$

It is to be noted that many field theories found useful today are globally
scale invariant. To see this, we first mention that the action is given by

\bigskip

$\int d^{4}x\mid g\mid^{1/2}L,$\ \ \ \ \ \ \ \ \ \ \ \ \ \ \ \ \ \ \ \ \ \
\ \ \ \ \ \ \ \ \ \ \ \ \ \ \ \ \ \ \ \ \ \ \ \ \ \ \ \ \ \ \ \ \ \ \ \ \
\ \

\bigskip

where $L$ is the Lagrangian density. By (3), the action is invariant under
global scale transformations provided that

\bigskip

$L^{\prime}=\Lambda^{-4}L.$ \ \ \ \ \ \ \ \ \ \ \ \ \ \ \ \ \ \ \ \ \ \ \
\ \ \ \ \ \ \ \ \ \ \ \ \ \ \ \ \ \ \ \ \ \ \ \ \ \ \ \ \ \ \ \ \ \ \ \ \
\ \ \ \ \ \ \ \ \ \ \ \ \ \ \ \ \ (4)

\bigskip

Consider the theory of a massless scalar field $\phi$ with a $\phi^{4}$
coupling. The Lagrangian of this theory is given by

\bigskip

$\dfrac{1}{2}g^{\mu\nu}\partial_{\mu}\phi\partial_{\nu}\phi-\lambda\phi^{4},$ 

$\ \ \ \ \ \ \ \ \ \ \ \ \ \ \ \ \ \ \ \ \ \ \ \ \ \ \ \ \ \ \ \ \ \ \ \ \
\ \ \ \ \ \ \ \ \ \ \ \ \ \ \ \ \ \ \ \ \ \ \ \ \ (5)$

where $\lambda$ is the quartic coupling constant. By (2), the Lagrangian
of
(5) satisfies (4) provided that

\bigskip

$\phi^{\prime}$=$\Lambda^{-1}\phi.$ \ \ \ \ \ \ \ \ \ \ \ \ \ \ \ \ \ \ \
\ \ \ \ \ \ \ \ \ \ \ \ \ \ \ \ \ \ \ \ \ \ \ \ \ \ \ \ \ \ \ \ \ \ \ \ \
\ \ \ \ \ \ \ \ \ \ \ \ \ \ \ \ \ \ \ \ \ \ \ \ (6)

\bigskip

Next consider the Lagrangion density of the electromagnetic field
$A_{\mu}$

\bigskip

$-\dfrac{1}{4}g^{\mu\rho}g^{\nu\sigma}F_{\mu\nu}F_{\rho\sigma},$ \ \ \ \ \
\ \ \ \ \ \ \ \ \ \ \ \ \ \ $\ \ \ \ \ \ \ \ \ \ \ \ \ \ \ \ \ \ \ \ \ \ \
\ \ \ \ \ \ \ \ \ \ \ \ \ \ \ \ \ \ \ \ \ \ \ (7)$

\bigskip

where

$\bigskip$

$F_{\mu\nu}=\partial_{\mu}A_{\nu}-\partial_{\nu}A_{\mu}.$

\bigskip

By (2), this Lagrangian density satisfies (4) provided that

$\bigskip$

$A_{\mu}^{\prime}=A_{\mu}.$ \ \ \ \ \ \ \ \ \ \ \ \ \ \ \ \ \ \ \ \ \ \ \
\ \ \ \ \ \ \ \ \ \ \ \ \ \ \ \ \ \ \ \ \ \ \ \ \ \ \ \ \ \ \ \ \ \ \ \ \
\ \ \ \ \ \ \ \ \ \ \ \ \ \ (8)

\bigskip

The Yang-Mills theory is also invariant under global scale
transformations.
This is because

\bigskip

$F_{\mu\nu}^{a}=\partial_{\mu}W_{\nu}^{a}-\partial_{\nu}W_{\mu}^{a}%
-gf^{abc}W_{\mu}^{b}W_{\nu}^{c}$

\bigskip

is invariant under global scale transformations provided that

\bigskip

$W_{\mu}^{\prime a}=W_{\mu}^{a}$, \ \ \ \ \ \ \ \ \ \ \ \ \ \ \ \ \ \ \ \
\ \ \ \ \ \ \ \ \ \ \ \ \ \ \ \ \ \ \ \ \ \ \ \ \ \ \ \ \ \ \ \ \ \ \ \ \
\ \ \ \ \ \ \ \ \ \ \ \ \ \ \ \ (9)

\bigskip

where $W_{\mu}^{a}$ is the Yang-Mills gauge meson, $g$ is a dimensionless
coupling constant, and $f^{abc}$ is the structure constant of the gauge
group.

The Lagrangian for the electron field $\Psi$ coupling with the
electromagetic
field and the gravitational field is

$\bigskip$

$\overset{-}{\Psi}i\gamma^{c}\varepsilon_{c}^{\mu}\left[  \partial_{\mu
}+ieA_{\mu}-\dfrac{1}{2}\sigma_{ab}\varepsilon^{b\nu}(\partial_{\mu
}\varepsilon_{\nu}^{a}-\Gamma_{\mu\nu}^{\rho}\varepsilon_{\rho}^{a})\right]
\Psi,$ $\ \ \ \ \ \ \ \ \ (10)$

\bigskip

where

$\bigskip$

$\Gamma_{\mu\nu}^{\rho}=\dfrac{g^{\rho\sigma}}{2}(\partial_{\mu}g_{\sigma\nu
}+\partial_{\nu}g_{\sigma\mu}-\partial_{\sigma}g_{\mu\nu}),$

\bigskip

$\sigma^{ab}=\dfrac{\gamma^{a}\gamma^{b}-\gamma^{b}\gamma^{a}}{4}.$

\bigskip

and $\varepsilon_{\mu}^{a}$ is the tetrad satisfying

\bigskip

$\eta_{ab}\varepsilon_{\mu}^{a}\varepsilon_{\nu}^{b}=g_{\mu\nu}.$

\bigskip

with $\eta_{ab}$ the metric tensor in the Minkowski space. Thus we have

\bigskip$\varepsilon_{\mu}^{\prime a}=\Lambda\varepsilon_{\mu}^{a}.$ $\ \
\ \ \ \ \ \ \ \ \ \ \ \ \ \ \ \ \ \ \ \ \ \ \ \ \ \ \ \ \ \ \ \ \ \ \ \ \
\ \ \ \ \ \ \ \ \ \ \ \ \ \ \ \ \ \ \ \ \ \ \ \ \ \ \ \ \ \ \ \ \ \ \ \
(11)$

\bigskip

 >From (2) and (11), we have

\bigskip

$\varepsilon^{\prime a\mu}=\Lambda^{-1}\varepsilon^{a\mu},$ $\ \ \ \ \ \ \
\ \ \ \ \ \ \ \ \ \ \ \ \ \ \ \ \ \ \ \ \ \ \ \ \ \ \ \ \ \ \ \ \ \ \ \ \
\ \ \ \ \ \ \ \ \ \ \ \ \ \ \ \ \ \ \ \ \ \ \ \ \ \ (12)$

\bigskip

where

\bigskip

$\varepsilon^{a\mu}=g^{\mu\nu}\varepsilon_{\nu}^{a}.$

\bigskip

It is then straightforward to find that the Lagrangian density (10)
satisfies
(4) provided that

\bigskip

$\Psi^{\prime}=\Lambda^{-3/2}\Psi.$ \ \ \ \ \ \ \ \ \ \ \ \ \ \ \ \ \ \ \
\ \ \ \ \ \ \ \ \ \ \ \ \ \ \ \ \ \ \ \ \ \ \ \ \ \ \ \ \ \ \ \ \ \ \ \ \
\ \ \ \ \ \ \ \ \ \ \ \ \ \ \ \ \ (13)

\bigskip

Needless to say, scale invariant Lagrangians do not contain any
dimensional
parameters, which may be generated by the mechanism of spontaneous
symmetry
breaking$^{3}.$

Let us see what happens when we demand the theory be scale invariant
locally,
i.e., when $\Lambda$ is a function of time and space. This means that we
require that the forms of all equations in the theory remain the same even
though we use different units of measurements at different space-time
points.
Again, this does not mean that there is a change of the distance between
two
points--it is just being measured in different units at different
space-time
points. With the well-known arguments employed to deduce the existence of
gauge fields, we find that all the theories above become locally scale
invariant provided that there exists a vector meson $S$ and we make in
these
Lagrangian the replacements

$\bigskip$

$\partial_{\mu}g_{\nu\rho}\rightarrow(\partial_{\mu}+2fS_{\mu})g_{\nu\rho},$
$\ \ \ \ \ \partial_{\mu}g^{\nu\rho}\rightarrow(\partial_{\mu}-2fS_{\mu
})g^{\nu\rho},$

$\partial_{\mu}\varepsilon_{\nu}^{a}\rightarrow(\partial_{\mu}+fS_{\mu
})\varepsilon_{\nu}^{a},$ \ $\ \ \ \ \ \ \ \ \ \partial_{\mu}\varepsilon
_{a}^{\nu}\rightarrow(\partial_{\mu}-fS_{\mu})\varepsilon_{a}^{\nu},$

$\partial_{\mu}\phi\rightarrow($ $\partial_{\mu}-fS_{\mu})\phi,$
$\ \ \ \ \ \ \ \ \ \ \ \ \partial_{\mu}\Psi\rightarrow(\partial_{\mu}%
-\dfrac{3}{2}fS_{\mu})\Psi.$ $\ \ \ \ \ \ \ \ \ \ \ (14)$

\bigskip

In the above, $f$ is a coupling constant which is the counterpart of $e$
in
quantum electrodynamics.

As an example, with the replacement (14), the Lagrangian of (5) becomes

$\bigskip$

$\dfrac{1}{2}g^{\mu\nu}(\partial_{\mu}-fS_{\mu})\phi(\partial_{\nu}-fS_{\nu
})\phi-\lambda\phi^{4}.$ $\ \ \ \ \ \ \ \ \ \ \ \ \ \ \ \ \ \ \ \ \ \ \ \
\ \ \ \ \ \ \ \ \ \ \ \ (15)$

$\ \ \ \ \ \ \ \ \ \ \ \ \ \ \ \ \ \ \ \ \ \ \ \ \ \ \ \ \ \ \ \ \ \ \ \ \
\ \ \ \ \ \ $

\bigskip The Lagrangian density above satisfies (4) with

$\bigskip$

$\phi^{\prime}=\Lambda^{-1}\phi$

\bigskip

$g^{\prime\mu\nu}=\Lambda^{-2}g^{\mu\nu},$

\bigskip

and

\bigskip

$S_{\mu}^{\prime}=S_{\mu}-\dfrac{1}{f}\partial_{\mu}\ln\Lambda,$
$\ \ \ \ \ \ \ \ \ \ \ \ \ \ \ \ \ \ \ \ \ \ \ \ \ \ \ \ \ \ \ \ \ \ \ \ \
\ \ \ \ \ \ \ \ \ \ \ \ \ \ \ (16)$
\ \ \

\bigskip

where $\Lambda$ is a function of space and time.

By (8) and (9), there is no need to alter the Lagrangian for a gauge
meson.
Thus gauge mesons do not couple with $S.$

Neither does a fermion, as the Lagrangian (10) is unchanged after the
replacement (14) is made. The proof is straightforward but requires a
little
algebra so we will give the details below. We note that, as we make the
replacement (14), the additional terms generated from

$\bigskip$

$-\dfrac{\gamma^{c}}{2}\sigma_{ab}\varepsilon^{b\nu}\varepsilon_{c}^{\mu
}(\partial_{\mu}\varepsilon_{\nu}^{a}-\Gamma_{\mu\nu}^{\rho}\varepsilon_{\rho
}^{a})$ $\ \ \ \ \ \ \ \ \ \ \ \ \ \ \ \ \ \ \ \ \ \ \ $

\bigskip

are

\bigskip

$-\dfrac{\gamma^{c}}{2}\sigma_{ab}\varepsilon^{b\nu}\varepsilon_{c}^{\mu}$
$f(-S_{\nu}\varepsilon_{\mu}^{a}+g^{\rho\sigma}S_{\sigma}g_{\mu\nu}%
\varepsilon_{\rho}^{a}).$

\bigskip

The expression above can be simplified into

\bigskip

$\dfrac{\gamma^{c}}{2}f(\sigma_{cb}\varepsilon^{b\nu}S_{\nu}-\sigma
_{ac}\varepsilon^{a\sigma}S_{\sigma})=f\gamma^{c}\sigma_{cb}\varepsilon^{b\nu
}S_{\nu}.$

$(17)$

\bigskip

Now

\bigskip

$\gamma^{c}\sigma_{cb}=\gamma^{c}\dfrac{\gamma_{c}\gamma_{b}-\gamma_{b}%
\gamma_{c}}{4}=\dfrac{3\gamma_{b}}{2}.$

\bigskip

\bigskip Thus the expression in (17) is equal to

\bigskip

$\dfrac{3}{2}f\gamma^{b}\varepsilon_{b}^{\nu}S_{\nu}.$

\bigskip

Also, the term generated from $\partial_{\mu}\Psi$ is

\bigskip

$-\dfrac{3}{2}fS_{\mu}\Psi.$

\bigskip

Therefore, as we make the replacement (14), the terms generated from the
Lagrangian (10) completely cancel one another. Thus we conclude that the
fermion does not couple with $S.$

With identical arguments, we may prove that quarks and leptons in the
standard
model do not couple with $S.$

That the vector meson $S$ does not couple to the electron avoids a
difficulty
which is distinct from the one raised by Einstein. As we know, the
Hamiltonian
of a non-relativistic electron in an electrostatic potential $eA_{0}$ is
$p^{2}/2m+eA_{0}.$ If the electron coupled to $S$ with the coupling
constant
$af$, this Hamiltonian would have, in addition, an imaginary potential
$iafS_{0}$. Such a Hamiltonian is not hermitian.

But the Lagrangian for a scalar field coupled to the $S$ field, given by
(15),
is real, and the Hamiltonian for such a theory is hermitian. \ The Higgs
meson, if it is found, will become the only known elementary particle,
beside
the graviton,  which interacts with the $S$ meson. If a Higgs meson is
created
in an experiment at a collider, it may emit an $S$ meson which manifests
its
presence by carrying away energy and momentum. However, the value of $f$
is
unknown. It appears that this value also determines the mass of the $S$
meson$^{3}$. If the $S$ mass is not exceedingly large, the value of $f$ \
may
be too small for this bremsstrahlung process to be observable.

Nuclei are formed as protons and neutrons attract one another with strong
interactions. Solids or liquids are formed as electrons and nuclei have
the
electromagnetic interaction. Since the $S$ meson has no such interactions,
it
cannot form the kinds of matter we see around us. Thus the $S$ particles
mostly move individually in space, with nothing to reveal their existence
other than through the gravitational field they generate. On the other
hand,
unlike electrons and nucleons, $S$ is a boson and does not satisfy the
exclusion principle. As a result, there is no limit to how many $S$
particles
may occupy the same spatial point at the same time. Therefore, when the
density of such particles become sufficiently high, it may be possible for
them to form a condensation the behavior of which is different from those
of
ordinary matters. In particular, it is translucent to light and likely
very
heavy. If these condensates exist, there may be translucent stars formed
by
this matter, or semi-translucent stars which are mixtures of the
translucent
matter and ordinary matters. It is interesting to ponder ways to observe
the
$S$ particles if they exist in space.

Finally, we comment on Einstein's objection to Weyl's theory on the ground
that the existence of the $S$ meson means that the length of a standard
rod
would depend on its history. Weyl continued to believe in 1955 that Einstein is
incorrect$^{4},$ and we concur with Weyl. The meaning of distance
is the same whether the $S$ meson exists or not.  While the state
of an electron does depend on its path, as in the Bohm-Aharonov
experiment performed by Chamber $^5$, this is consistent with the
principle of quantum mechanics.

\bigskip

Acknowledgements

\bigskip

I like to thank Dr. D. Margetis and Professor M.Van Putten for helpful
discussions.

\bigskip

\bigskip

References

1.H. Weyl, Z.Phys.56,330 (1929).

2.See, for example, J.T.Wheeler, J. Math.Phys. 39(1), 299,1998, and the
references quoted in this paper.

3.H.Cheng, Phys. Rev. Letters 61, 2182 (1988).

4. See Author's reply at the end of the paper by H. Weyl, Gravitation and
Electricity, in the book entitled ''The Dawning of Gauge Theory'' by
Lochlainn
O'Raifeartaigh, Princeton Series in Physics, Princeton University Press,
Princeton, New Jersey.

5.R.G.Chambers, Phys.Rev.Lett. 5.3 (1960).

\bigskip
\end{document}